\documentclass[sigconf, nonacm]{acmart}
\usepackage{tabularray}
\usepackage{soul}
\UseTblrLibrary{siunitx, varwidth}
\usepackage{etoolbox}
\newrobustcmd\B{\DeclareFontSeriesDefault[rm]{bf}{b}\bfseries} 

\AtBeginDocument{%
  \providecommand\BibTeX{{%
    \normalfont B\kern-0.5em{\scshape i\kern-0.25em b}\kern-0.8em\TeX}}}

\setcopyright{acmcopyright}
\copyrightyear{2023}
\acmYear{2023}
\acmDOI{XXXXXXX.XXXXXXX}

\acmConference[CIKM '23]{The 32nd ACM International
Conference on Information and Knowledge Management}{June 03--05,
  2018}{Woodstock, NY}
\acmPrice{15.00}
\acmISBN{978-1-4503-XXXX-X/18/06}

\begin{document}


\title{ClusterSeq: Enhancing Sequential Recommender Systems with Clustering based Meta-Learning}


\author{Mohammmadmahdi Maheri}
\affiliation{%
  \institution{Imperial College London}
  \streetaddress{}
  \city{London}
  \country{United Kingdom}}
\email{m.maheri23@imperial.ac.uk}

\author{Reza Abdollahzadeh}
\affiliation{%
  \institution{Sharif University of Technology}
  \city{Tehran}
  \country{Iran}
}
\email{re.abd@student.sharif.edu}

\author{Bardia Mohammadi}
\affiliation{%
  \institution{Sharif University of Technology}
  \city{Tehran}
  \country{Iran}
}
\email{bardia.mohammadi@sharif.edu}

\author{Mina Rafiei}
\affiliation{%
  \institution{Sharif University of Technology}
  \city{Tehran}
  \country{Iran}
}
\email{m.rafiei@sharif.edu}

\author{Jafar Habibi}
\affiliation{%
  \institution{Sharif University of Technology}
  \city{Tehran}
  \country{Iran}
}
\email{jhabibi@sharif.edu}

\author{Hamid R. Rabiee}
\affiliation{%
  \institution{Sharif University of Technology}
  \city{Tehran}
  \country{Iran}
}
\email{rabiee@sharif.edu}

\renewcommand{\shortauthors}{Maheri, et al.}
\newcommand{\rafiei}[1]{\textcolor{red}{rafiei: #1}}



\begin{abstract}
In practical scenarios, the effectiveness of sequential recommendation systems is hindered by the user cold-start problem, which arises due to limited interactions for accurately determining user preferences. Previous studies have attempted to address this issue by combining meta-learning with user and item-side information. However, these approaches face inherent challenges in modeling user preference dynamics, particularly for "minor users" who exhibit distinct preferences compared to more common or "major users." To overcome these limitations, we present a novel approach called ClusterSeq, a Meta-Learning Clustering-Based Sequential Recommender System. ClusterSeq leverages dynamic information in the user sequence to enhance item prediction accuracy, even in the absence of side information. This model preserves the preferences of minor users without being overshadowed by major users, and it capitalizes on the collective knowledge of users within the same cluster. Extensive experiments conducted on various benchmark datasets validate the effectiveness of ClusterSeq. Empirical results consistently demonstrate that ClusterSeq outperforms several state-of-the-art meta-learning recommenders. Notably, compared to existing meta-learning methods, our proposed approach achieves a substantial improvement of 16-39\% in Mean Reciprocal Rank (MRR).

\end{abstract}






\maketitle

\section{Introduction}
Recommender systems achieve much more attention today due to the need for personalization in a wide range of applications. They aim to consider the user preferences to suggest the most suitable item for the user.

One common classification of the recommender systems is based on their way of extracting users' preferences. It classifies the recommender systems as collaborative
filter-based, content-based, or hybrid. Through collaborative filtering, numerous users' preferences are collected to estimate the user's response. To predict and recommend new items to a user, the similarity of the user's history is compared with existing users, and the predictions are made based on their existing ratings. This approach's effectiveness heavily depends on the presence of substantial prior interactions between users and items. Therefore, in the cold-start situation where users and items have a sparse history of interactions, they could not perform well.

To solve the cold starting problem, content-based systems were introduced \cite{mooney2000content,narducci2016concept}. They use the side information of the users and items to make recommendations. They rely on the similarity of a user with items. Therefore, they cannot consider the user history of items to calculate user preferences. Also, they cannot be efficient if the side information is unavailable for various reasons, including privacy. Although hybrid systems use collaboration and content information simultaneously, they do not adequately solve the mentioned challenges.

Sequential recommendation also plays a crucial role in a real-world application. The goal is to extract user preferences based on the sequence of user's interactions to predict more possible items that the user would have an interaction with. A significant challenge can be long-tailed interaction data in some real-world scenarios due to limited interactions by new users \cite{yin2020learning}. Several sequential recommendation problems can benefit from transformer-based approaches, including SASRec \cite{kang2018self} and BERT4Rec \cite{sun2019bert4rec}, which consistently capture dynamic behaviors. However, in real-world scenarios, the sequence length of the cold users is not long enough for these methods to be effective \cite{liu2021augmenting}.

Cold-start can deteriorate recommender system performance and. The majority of existing cold-start methods rely on auxiliary information or knowledge from other domains. In a cold-start sequential recommendation, side information is absent, and the goal is to model dynamic user behavior even in short sequences. Sharing knowledge among users could also improve performance, especially for cold-users and cold-items. In response, meta-learning-based research has generated promising results against the cold-start problem in recent years \cite{lee2019melu}. Meta-learning algorithms could improve user recommendations by utilizing the meta-knowledge extracted from the users to enhance new user recommendations. Although meta-learning has many advantages, it has one fundamental disadvantage when used in recommender systems. It is rooted in the generalization meta-learning algorithms aim for. Since users have different preferences, there is a possibility that each user might have a similar preference, like most users, or a preference that is less popular. When designing a recommender system, it is important to take into account the needs of both minor and major users. In contrast, meta-learning approaches bias the model parameters by the major users and get stuck in local minima.

In this paper, we propose a meta-learning-based sequential recommender system named ClusterReq that addresses the preceding problems. It utilizes the Model-Agnostic Meta-Learning algorithm (MAML) \cite {finn2017model}, to share knowledge among users and adapt quickly to new users with limited transactions. Hence, a personalized recommendation is provided for each user based on their item transaction history. Then, a robustly designed clustering approach modulates the network parameters to prevent local minima made by the major user group. As a result, the proposed model can take into account transitional dynamic preferences and changes in preferences even in short sequences. The contributions of this work can be summarized as follows: 


\begin{itemize}

\item  Utilizing a meta-learning approach to design an architecture to address sequential recommendation without side information and allow fast adaptation for users who start cold.

\item Solving the challenge of meta-learning bias towards major users in recommendation systems by clustering users and modulating their parameters to avoid local minimums.

\item Achieving 16-39\% improvement in MRR, on three real-world benchmarks compared with the state-of-the-art.
\end{itemize}








\section{Related Works}

\textbf{Sequential Recommendation.}
The sequential recommender systems model user preferences based on the sequence provided for each individual. Users' preferences have been modeled in these systems using a variety of deep learning architectures in recent years. Users were embedded in a transition space in TransRec \cite{he2017translation} so that each user could be considered as a vector connecting sequence items. More recently, convolutional neural networks \cite{yuan2019simple,tang2018personalized} and recurrent neural networks \cite{hidasi2015session,hidasi2018recurrent} were used to find appropriate user embedding that takes into account the order of sequence items. Moreover, transformers solve the sequential recommendation problem more efficiently by prioritizing the entire sequence without forgetting earlier items in the sequence \cite{kang2018self,sun2019bert4rec}. Finally, graph neural networks were adopted to solve the problem by utilizing graphical information \cite{wang2021session,wu2019session}. The above-mentioned works have contributed to the effective extraction of more information from an interaction sequence. In contrast, they could not be effective if the user behavior sequence is short in a cold-start scenario.


\textbf{Meta-learning for Recommendation.}
It is possible to use meta-learning in few-shot learning due to its ability to adapt rapidly with only a small number of samples. Previous research has used MAML-based models to solve the user-cold start problem \cite{wei2020fast,lee2019melu}. To capture meta-path semantics, MetaHIN \cite{lu2020meta} integrates heterogeneous information networks (HINs) into MAML. To adapt well to different tasks and avoid local optimum, MAMO proposes different task-specific adaptation strategies \cite{dong2020mamo}. It introduces two memory arrays: a feature-specific memory for initializing shared parameters and a task-specific memory for guiding the model. Recently, an attempt was made by TaNP \cite{lin2021task} to solve MAML problems such as model sensitivity and local optima. It customizes global knowledge to task-related decoder parameters for estimating user preferences. To capture task dependency more effectively, PAML \cite{wang2021preference} considers user relations with defined palindrome paths among them. Therefore, model parameters can be more precisely tailored to capture diverse paths and interactions. More recently, \cite{pang2022pnmta} was developed to alleviate the problem of limitations of representing users imposed by the particular task setting of MAML-based recommender systems. Firstly, a pre-trained model was obtained with a non-meta-learning method. After that, an encoder modulator corrects the prior parameters for the meta-learning task. The authors in \cite{wang2021sequential} examined the problem of sequential recommendation and extracting transitional patterns among users' sequences without needing side information of users and items.

Several previous works \cite{dong2020mamo,lin2021task,wang2021preference,pang2022pnmta,song2021cbml} have attempted to address MAML's potential problems in recommendation systems. The problem relates specifically to the elimination of preferences belonging to minor users. These users are less prevalent in the training data than the majority of users. However, \cite{dong2020mamo,lin2021task,song2021cbml} primarily relies on k-means-based clustering, which is inappropriate for many scenarios where users could be divided into primary and minor groups. To cluster users, they also require auxiliary information from users and items. In many scenarios, such data does not exist. Lastly, clustering users based on considering interacting items independently could not be effective, especially in cold-start scenarios in which a limited number of items are available for a user.

The method in \cite{wang2021preference,pang2022pnmta} is based on modulating prior knowledge based on user embeddings. However, the modulation would be biased toward major users such that majors would collapse modulation for other minor clusters. In addition, they need side information like explicit friends in \cite{wang2021preference} to embed users properly. If limited data is available about explicit friends, joint inference methods could estimate the missing information \cite{ramezani2023joint}, but without that information modulation will not benefit learning.  Also, embedding users without considering the sequence of items or users' similarity would not be practical for sequential recommendations. It is worth noting that CBML \cite{song2021cbml} relies on computing gradients for users to cluster them. This can result in higher computational costs than other models. Additionally, CBML may require multiple support adaptation steps in more complex scenarios to ensure effective performance.

\section{The proposed model}
In this section, the proposed model will be discussed in detail by describing its ability to address the previously mentioned challenges and their solutions. Specifically, the model is designed to satisfy the following goals: 1) Sharing knowledge among different users to improve performance on cold-start users, 2) Capturing short-range and long-range dynamics in users' preferences, 3) Optimizing the model to perform accurately on cold-start users, and 4) Avoiding local minima in shared parameters and considering major and minor preferences, simultaneously.

\subsection{Problem Setup}
We assume there exists a set of users $U$ divided between training and testing ($U_{train} \cap U_{test} = \emptyset$), and items $I$. The model's goal is to predict the next-item preference score of all items and recommend top-N items by inputting a short chronological order sequence of a user's preference $Seq_u = (i_{u,1},i_{u,2},...,i_{u,k-1})$. Note that the items and the users have no auxiliary information, and embedding is calculated only by their ID. 

To mimic the cold-start scenario, the users with a limited number of transactions $(K)$ will exist in the $U_{test}$. Based on the proposed optimization strategy and architecture, the model's parameters will adapt quickly to the user's short available transactions. The model's performance for predicting the next item of test users will be reported.  

\subsection{Proposed Architecture}
The following sections provide detailed information about each module's components and its significance. We will first discuss the meta-learning approach to overcoming the few-shot problem in section \ref{subsection:few-shot}. In section \ref{subsection:model_overview}, we will provide an overview of the proposed architecture. The details of the sub-modules will be described in sections \ref{subsection:dynamic_transition} and \ref{subsection:clustering_module}. Finally, we describe how to optimize the model's parameters in the meta-learning setting in section \ref{subsection:optimization}.

\subsubsection{Few-shot Recommendation}
\label{subsection:few-shot}
Meta-learning could transfer knowledge from data-rich users to cold users. By considering each user as a separate task, meta-learning extracts common knowledge among users. Therefore, a cold-start user needs much less data to converge to its optimal parameters, and user preferences could be detected with a few-shot approach. When using meta-learning, you must define support and query sets. 

The support set of a user is used to adapt shared knowledge (parameters) to the user. Thus, the user's personalized parameter will be adapted to the support set. Afterward, the user's query set is used to evaluate the adaptation. We assume users with a constant (K) sequence length as:  

\[
I_{1}
\to
I_{2}
\to
...
\to
I_{k-1}
\to
I_{k}
\]

We define each user $u$ as a task $\tau_{u}$, and its sequence is divided into support and query sets. First, $k-1$ items in a sequence are considered support-set, and the last item of the sequence is in query-set. Meta-learning tries to predict the query set by adapting to the previous items in the support set. 

The approach to sampling training data also needs to be designed. Users usually have a wide range of history lengths that should be used efficiently to extract sequential patterns. Users' history of transactions should also be appropriately utilized, even in a short user sequence, to improve performance for cold-start users. Sampling the training data should mimic cold-start users. Therefore, it only samples a short sequence of transactions for each user. First, a user is sampled from $U_{train}$ to construct a meta-training task. All user interactions will then be limited to a sequence of length $K$. In this way, the training is more like a test scenario in which test users will be considered.

\subsubsection{Model Overview}
\label{subsection:model_overview}

\begin{figure}[ht]
  \centering
  \includegraphics[width=\linewidth]{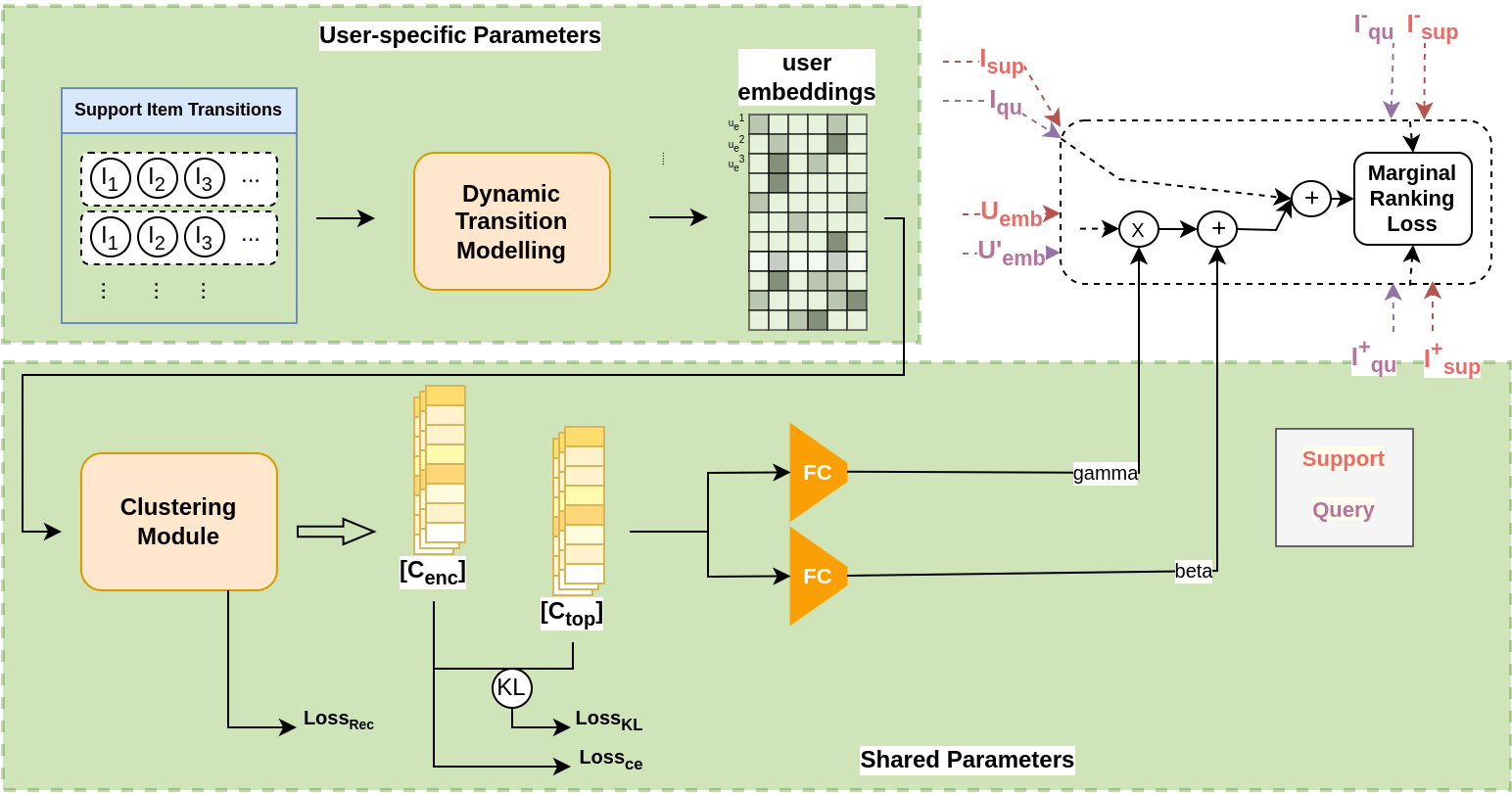}
  \caption{The overall architecture of the model, including the clustering module and dynamic transition embedder}
  \Description{}
  \label{fig:overallarchitecture}
\end{figure}

In terms of capturing both short-range and long-range dynamics of users' preferences, and considering all major and minor groups of users, we designed a dynamic transition modeling and clustering module. This will be explained in the following sections \ref{subsection:dynamic_transition} and \ref{subsection:clustering_module} respectively.

Figure \ref{fig:overallarchitecture} shows the overall view of our proposed model. First, the Dynamic Transition model converts item transitions to user embeddings. Then, using a clustering module, we calculate two soft clustering assignments. These assignments are used to calculate loss values and condition user embeddings for the next item prediction. Finally, we score positive and negative item predictions and propagate it backwards on the marginal ranking loss.

\subsubsection{Dynamic Transition Modeling}
\label{subsection:dynamic_transition}
\begin{figure}[ht]
  \centering
  \includegraphics[width=\linewidth]{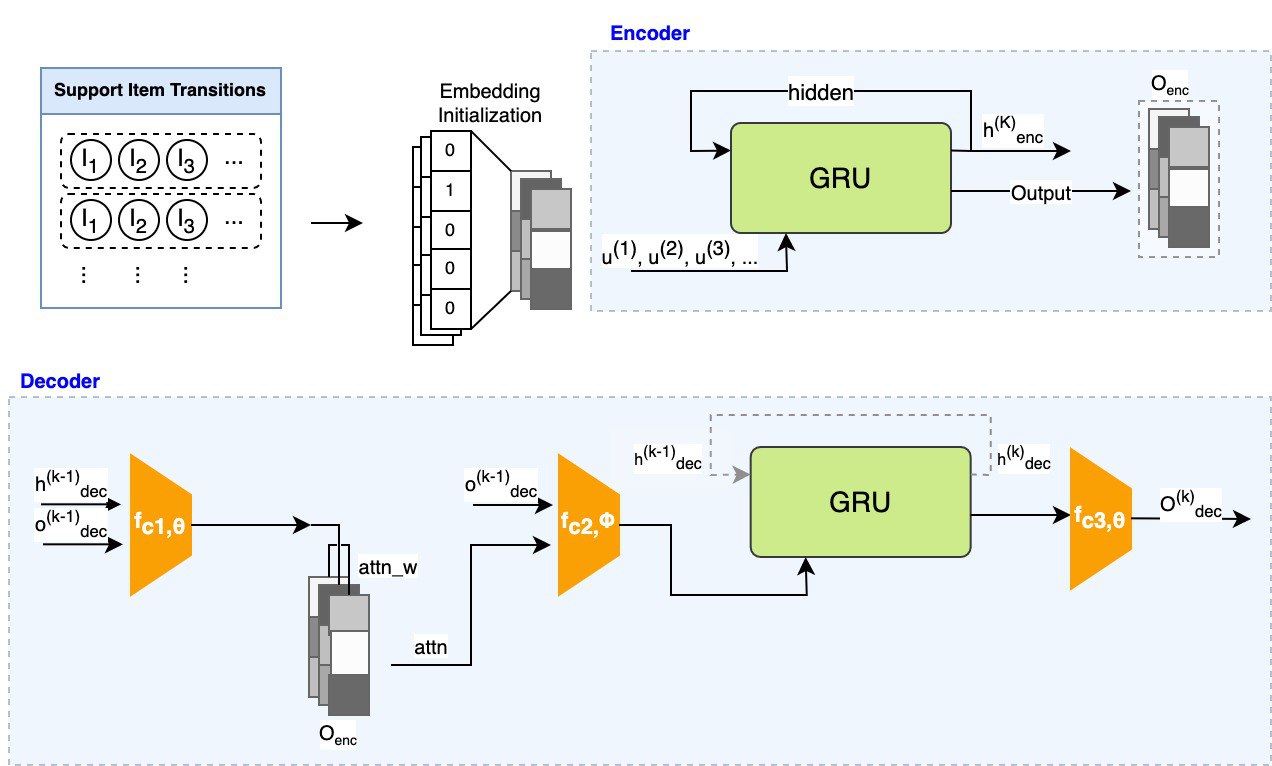}
  \caption{Architecture of dynamic transition modeling}
  \Description{}
  \label{fig:modelarchitecture}
\end{figure}

Considering sequential patterns and extracting dynamic patterns needs a particular architecture in few-shot settings. Also, the model needs to detect temporal and long-term user preference changes. As shown in Figure \ref{fig:modelarchitecture}, attention-based recurrent neural network architecture is proposed to satisfy these goals.

This architecture consists of an encoder and a decoder incorporating an attention mechanism. In the first stage of the process, the user transition sequence $u_i$ is fed to the encoder, which is essentially a Gated Recurrent Unit (GRU) gate \cite{chung2014empirical}:
\begin{equation}
\begin{aligned}
    \label{eq:dyn_trans_enc}
    o_{i, enc}^{(k)} &= f_{enc}(u_i^{(k-1)}, h_{i, enc}^{(k-1)}) \\
    h_{i, enc}^{(k)} &= h_{enc}(u_i^{(k-1)}, h_{i, enc}^{(k-1)})
\end{aligned}
\end{equation}
$f_{enc}$ and $h_{enc}$ are GRU functions to generate its output and hidden vector, respectively. We consider $O_{i, enc}\in \mathbb{R}^{K\times D} = \{o_{i, enc}^{(k)}\}$ the primary context vector of this sequence. $h_{i, enc}^{(0)}$ is initialized by a zero vector. \\
As for the decoder, each sequence item passes through the network in different iterations. We use the decoder output vector $o_{i, dec}^{(k-1)}$ and hidden value $h_{i, dec}^{(k-1)}$ of the previous iteration to apply attention to the context vector for the current one:
\begin{equation}
\begin{aligned}
    \label{eq:dyn_trans_dec_attn}
    attn\_{w_i}^{(k)} &= softmax(fc_1(o_{i, dec}^{(k-1)};h_{i, dec}^{(k-1)})) \\
    attn_i^{(k)} &= attn_{w_i}^{(k)} \cdot O_{i, enc}
\end{aligned}
\end{equation}

Here, the semicolon sign $;$ shows the concatenation of vectors, $fc_1 \colon \mathbb{R}^{2D} \mapsto \mathbb{R}^{K}$ is a one-layer fully connected network and $attn\_w_i^{(k)} \in \mathbb{R}^{K}$ indicates attentional weights to be applied to the context vector concerning the previous $k-1$ items. The following will calculate the decoder's output and hidden vectors:

\begin{equation}
\begin{aligned}
    \label{eq:dyn_trans_dec_gru}
    X_i^{k} &= relu(fc_2(o_{i, dec}^{(k-1)};attn_i^{(k)})) \\ 
    o_{i, dec}^{(k)} &= softmax(fc_3(f_{dec}(X_i^{k}, h_{i, dec}^{(k-1)}))) \\
    h_{i, dec}^{(k)} &= h_{dec}(h_{i, dec}^{(k-1)})
\end{aligned}
\end{equation}

We consider the last encoder hidden vector $h_{i, enc}^{(K)}$ as $h_{i, dec}^{(0)}$ and a zero vector as $o_{i, dec}^{(0)}$. Also, $fc_2 \colon \mathbb{R}^{2D} \mapsto \mathbb{R}^{D}$ and $fc_3 \colon \mathbb{R}^{D} \mapsto \mathbb{R}^{D}$ are one-layer fully connected networks and $f_{dec}$ and $h_{dec}$ are the output and hidden functions of the GRU gate.



\subsubsection{Clustering Module}
\label{subsection:clustering_module}
\begin{figure}[ht]
  \centering  
  \includegraphics[width=\linewidth]{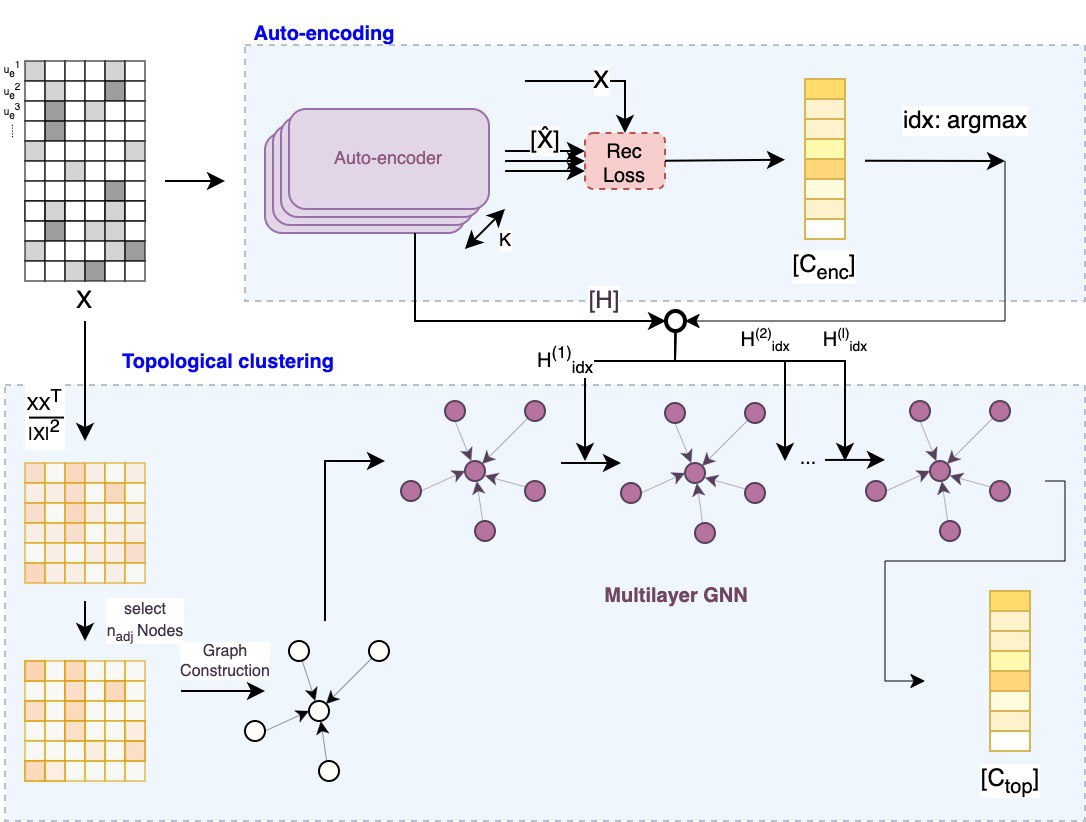}
  \caption{Architecture of clustering module}
  \label{fig:cluster_fig}
  \Description{}
\end{figure}

Our goal should be to consider both major and minor preferences sufficiently, and avoid major users distorting the model parameters as shown in Figure \ref{fig:cluster_fig}. In order to accomplish this, a clustering model is developed. Based on the fact that only implicit interactions between users are available, the clustering must be based solely on these interactions. Clustering, however, would be beneficial only if the major cluster did not disrupt clustering by attracting all the points to its center.

To address the problem of collapsing by major users, the proposed module is inspired by \cite{opochinsky2020k}. More specifically, it consists of a Graph Convolutional Network (GCN) to capture topological information from users' interaction data, combined with a K Auto Encoder (KAE) clustering module. The GCN network determines the topological clusters, and the auto-encoder chooses the encoding cluster based on the auto-encoded number. In order to accomplish this, we must first reconstruct user embeddings using $M$ randomly initialized autoencoders. Each autoencoder reconstructs the user embedding with the corresponding reconstruction loss. Therefore, each user's encoding cluster assignment is based on its reconstruction loss. The encoding clustering assignments are calculated according to equation \ref{eq:kae_cluster_assignment}.
\begin{equation}  
    \label{eq:kae_cluster_assignment}
    c_{enc}^{u} = softmax([-d(e_u,\hat{e}_u (i))])
\end{equation}
$c_u$ is the assignment of user $u$. In addition, $e_u$ and $ \hat{e}_u (i)$ are embeddings of the user and their reconstructions by the $i$th autoencoder, respectively, and $d$ is the $L2$ distance norm. The reconstruction loss of the autoencoders should also be calculated. The loss is calculated based on \ref{eq:kae_loss}:

\begin{equation}
    \label{eq:kae_loss}
    \mathcal{L}_{rec} = L(\theta_1, ... , \theta_M) = min_i d(e_u,\hat{e}_u (i))
\end{equation}
Where $\theta_j$ is the parameters of the $j$th autoencoder.

Then, we construct a users' relation graph $G$ using user embedding cosine similarity:

\begin{equation}
\begin{aligned}
    \label{eq:user_relation_graph}
    \forall{u_i \in{U_{train}}}, \forall{(u_i, u_j)\in{G}}; \\ \{u_j\} = argmax_{A'\subset{U_{train}-{u_i}}, |A'|=n_{adj}} \\ {\frac{u_i.u_j}{|u_i||u_j|} + \sigma \sum_{i_1, i_2 \in shared(u_i,u_j)}}{\frac{i_1.i_2}{|i_1||i_2|})}
\end{aligned}
\end{equation}


Here, for each user embedding $u_i$ we select $n_{adj}$ other user embeddings with the maximum user and item similarity values with $u_i$ to construct a graph $G$. As for the user similarity value, we compute the cosine similarity between user embeddings. As for item similarities, we calculate the sum of the cosine similarity values of all shared items between $u_i$ and $u_j$ ($shared(u_i,u_j)$). Here, $\sigma$ is a user-item balancing hyperparameter.

We assume the autoencoder with minimum reconstruction loss for user $u$ has $L = L_{enc} + L_{dec}$ layers and $H_u^{(l)}$ is the representation learned by layer $l$ of this autoencoder.
Here, $H_u^{(0)}$ is the user dynamic transition embedding. 
Corresponding to each encoder layer, there is a GCN layer ($L_{enc}$ layers in total) which we show by $Z^{(l)}$ and is calculated sequentially for user $u$ as follows:

\begin{equation}
\begin{aligned}
    \label{eq:gcn_update}
    Z_u'(l) &= (1-\epsilon)Z_u^{(l-1)} + \epsilon H_u^{(l-1)} \\
    Z_u^{(l)} &= \phi{(AZ_u'^{(l)}W_l)}
\end{aligned}
\end{equation}
Taking $B$ as the batch size, $D_{l, in}$ and $D_{l, out}$ as the in and out dimensions of $l$th layer of autoencoders,
 $A \in \mathbb{R}^{B\times B}$ is the normalized adjacency matrix of graph $G$ and $W_l \in \mathbb{R}^{D_{l, in}\times D_{l, out}}$ 
 is the weight matrix of $l$th layer. Using the GCN output, we calculate the topological clustering assignment of $u$:

 \begin{equation}
\begin{aligned}
    \label{eq:top_clustering}
    c_{top}^u = softmax(Z_u^{(L_{enc})}) \\
\end{aligned}
\end{equation} 

We optimize clustering assignment $c_{enc}^u = [c_{i,j}]$ by closing representations to cluster centers and making harder assignments using KL divergence loss: 

\begin{equation}
\begin{aligned}
    \label{eq:top_clustering_modified}
    c_{enc\_mod}^u = [c'_{i,j}]; c'_{i,j} = \frac{c_{i,j}^2 / f_{i,j}}{\sum_{j'}{c_{i,j'}/f_{i,j'}}} \\
    \mathcal{L}_{mod} = KL(c_{top}^u || c_{top_{mod}}^u)
\end{aligned}
\end{equation} 
where $f_{i,j} = \sum_{i}{c_{i,j}}$. This clustering assignment is used to supervise $c_{enc}^u$ using KL divergence loss:

\begin{equation}
\begin{aligned}
    \label{eq:kld_top_enc_clustering}
    \mathcal{L}_{combo} = KL(c_{enc\_mod}^u || c_{top}^u)
\end{aligned}
\end{equation} 

The overall clustering loss is calculated as the sum of all previously mentioned losses:

\begin{equation}
\begin{aligned}
    \label{eq:clustering_loss}
    \mathcal{L}_{CM} =
    \mathcal{L}_{rec} + \mathcal{L}_{mod} + \mathcal{L}_{combo}
\end{aligned}
\end{equation} 

The user embedding $u$ is then conditioned on the encoding clustering assignment $c_{enc}^u$ as shown in 
\ref{eq:clustering_conditioning}:

\begin{equation}
\begin{aligned}
    \label{eq:clustering_conditioning}
    u' = f_{gamma}(c_{enc_{mod}}^u).u + f_{beta}(c_{enc_{mod}}^u)
\end{aligned}
\end{equation} 
where $f_{gamma}$ and $f_{beta}$ are one-layer fully connected networks and $.$ is element-wise multiplication.



\subsubsection{Optimization and Fast Adaptation}
\label{subsection:optimization}
The proposed model needs proper optimization to transfer extracted knowledge to new users and adapt personalized parameters based on their few transactions. Specifically, some initialized parameters will be adjusted to support a user's set, to produce personalized parameters. In contrast, other parameters are not personalized for each user and shared. We denote the parameters that will not adapt to support set and are shared among tasks (users) with $\omega$. Other parameters are denoted by $\Phi$.   
It would be better to share the clustering module parameters among all tasks since this module detects similar tasks, as shown in Figure \ref{fig:overallarchitecture}.
 In contrast, the user's preference extractor will be adapted by each user, so its parameters are in the personalized parameter set $\Phi$. In addition, all model parameters are denoted by $\theta$ $(\theta = \omega \cup \Phi)$. Therefore, user-specific parameters could be adapted to the support set by equation \ref{eq:user_specific_update}:

\begin{equation}
    \label{eq:user_specific_update}
    \omega'_{n} = \omega - \alpha \nabla_{\omega} \mathcal{L}_{S_n}
\end{equation}

\noindent In which $\alpha$ is the task learning rate (local learning rate) and $\mathcal{L}_{S_n}$ are the losses calculated on the support set data and $\nabla_{\omega} \mathcal{L}_{S_n}$ is its corresponding gradient. More specifically, the support loss is based on ranking margin loss. This is the difference between the score of the positive and negative items' scores. The positive sample is the item in the actual sequence of a user, and the negative samples are items that are not a member of the user's actual sequence and are sampled randomly. The support loss $\mathcal{L}_{S_n}$ is calculated based on the equation \ref{eq:support_loss}:

\begin{equation}
\begin{aligned}
    \label{eq:support_loss}
    \mathcal{L}_{S_n} = \sum_{i=3}^{K-1} max(0,\lambda + s(I_{i-1} \xrightarrow[]{} I_i) - s(I_{i-1} \xrightarrow[]{} I_i'))\\
    s(I_{i-1} \xrightarrow[]{} I_i) = \lVert 
    f( \{I_0,...,I_{i-1} \} , \theta)  - I_i
    \rVert _{2}
\end{aligned}
\end{equation}

$\lambda$ represents the margin value, and $I_k'$ represents the negative sample, an item that has never interacted with the user $u_n$. Also, $s(.)$ is the function that calculates the score of the predicted item compared with the ground truth ($I_i$) on the given parameters of the neural network $\theta$. The function $f(.)$ symbolizes the model's output based on the user's history sequence.

The query loss will be calculated based on the $\omega'_{n}$ on the query set $Q_n$. Specifically, the query loss is calculated by \ref{eq:query_loss}:

\begin{equation}
    \label{eq:query_loss}
    \mathcal{L}_{Q_n} = 
    max(0,\lambda + s(I_{K-1} \xrightarrow[]{} I_K) - s(I_{K-1} \xrightarrow[]{} I_K'))
\end{equation}

Based on the meta-learning goals, the learning process tries to find the initial parameters for $\theta$, which reduces the overall loss of the query sets of all tasks. More specifically, the overall loss is calculated by \ref{eq:overal_loss}:


\begin{equation}
    \label{eq:overal_loss}
    \theta = \min_{\theta} \sum_{\tau_n \in p(\tau)}\mathcal{L}_{Q_n}(\omega'_n,\phi) + 
    \mathcal{L}_{CM_n}
\end{equation}

$\mathcal{L}_{CM_n}$ is the loss corresponding to the clustering module for $u_n$, which was defined in \ref{eq:clustering_loss}.
As the equation defines, meta-training tasks come from a distribution, and we only have a random sample set in the training phase. So by minimizing their query loss on the adopted parameters, the algorithm will converge to a proper initialization parameter for the tasks. The query loss is calculated based on the parameters adopted on each user's training support set. To solve the equation, the Stochastic Gradient Descent (SGD) is used in \ref{eq:sgd}:

\begin{equation}
    \label{eq:sgd}
    \theta \xleftarrow{} \theta - \beta \nabla_{\theta}  \sum_{\tau_n \in p(\tau)} \mathcal{L}_{Q_n}(\omega'_n,\phi)
\end{equation}

\noindent in which $\beta$ is the meta-learning rate.

The test phase is almost similar to the training phase. For each cold-user not seen in training, the task-specific parameters are adapted to its support set as in \ref{eq:user_specific_update}. Secondly, the score of the target item in the query set will be compared with 100 random negative samples to calculate the evaluation metrics.





\section{Experiment}

\subsection{Experimental Setup}


\begin{table}[]
\caption{  Dataset Statistics }
\label{table:statistics}
\begin{tabular}{cccc}
\hline
Dataset     & Users & Items  & Avg. Length of Sequence \\ \hline
Electronics & 29710 & 20712  & 13.51                   \\
Movies      & 199435 & 155527  & 10.87                   \\
Beauty      & 82659 & 124859 & 6.96                   \\ \hline
\end{tabular}
\end{table}

\subsubsection{Datasets}

From Amazon, we have adopted three widely used real-world datasets, which are shown in Table \ref{table:statistics}. The Electronics dataset is derived from the public Amazon review dataset. This includes reviews of Amazon products belonging to the "Electronics" category from May 1996 to July 2014. Both The Movies and The Beauty are drawn from the same "Movie" and "Beauty" Amazon review categories. User reviews are treated as an interaction between them. These interactions are treated equally on all items. The $K$ parameter specifies the minimum number of transactions a user must keep. In addition, we delete users with fewer interactions in the system. We sort the data in order of the first transaction time, user ID, and transaction time. We then assign a new label to the items and users according to their appearance time, so that the first user is one, the second user is two, etc. Lastly, we will separate the test data from the test items. We will only keep the test data that contains interactions that use items from the test items in the test data. For each user node in the test and validation sets, we take each observed edge as a positive sample of the user. We then randomly select 100 items that did not interact with the current user as negative samples. Then based on the rank of the positive sample's score among negative samples, evaluation metrics will be calculated as in

\cite{wang2021sequential,wei2020fast,kang2018self}.

\subsubsection{Baselines}

We compare the proposed model with the following methods: 

(i) Sequential recommendation baselines utilize different methods to capture the sequential patterns in the interaction sequences of users:

\begin{itemize}
    \item SASRec: presents a self-attentive sequential recommendation model that utilizes Gated Recurrent Units, a simple convolutional generative network, and a self-attention mechanism to capture sequential patterns in user behavior and improve recommendation accuracy. The model is trained using a modified BPR loss function. 

    \item BERT4Rec: proposes a novel recommendation model that uses the BERT architecture to capture sequential patterns in user behavior and improve recommendation accuracy. The model is trained using the bi-directional transformer to extract sequential patterns, outperforming other state-of-the-art models in accuracy and robustness to cold-start and long-tail item problems. The paper acknowledges some BERT4Rec limitations, such as its computational complexity and data requirements. However, it argues that the model's benefits justify the additional computational resources. 

\end{itemize}

(ii) Cold-start baselines include methods that provide accurate recommendations for customers with limited information. We modify these cold-start baselines to fit the case without auxiliary information. To deal with this issue, in the no side-information setting, for the datasets, we convert them into implicit recommendations by setting rated items to 1 and others to 0, and we utilize the Marginal Ranking loss function, which is the same as in our model, as we make implicit recommendations for binary signals. We just use the ID embedding of users and items as a feature (some methods like NGCF and LightGCN use this kind of embedding). In the training phase of recommendations, we sample data from the user and corresponding positive and negative items to calculate the loss at each step.

\begin{itemize}
    \item MeLU: Resolve the cold-start problem faced by existing recommender systems. The MeLU method uses meta-learning to estimate new users' preferences based on items they have consumed in the past. Moreover, the system provides a strategy to select evidence candidates to estimate customized preferences. It is shown that MeLU has a lower mean absolute error than two comparative models when tested on two benchmark datasets. In addition, the evidence selection strategy is tested in a user study. It aims to overcome the limitations of previous recommendation studies. These studies provided poor recommendations for users who consumed few items and inadequate evidence for candidates to identify user preferences.

    \item MetaTL: For cold-start users with minimal logged interactions, capturing sequential patterns of users for sequential recommenders is challenging. Models with limited interactions lose their predictive power due to difficulties in learning sequential patterns. Using meta-learning, the method proposes an innovative MetaTL framework that models users' transition patterns. A translation-based architecture extracts dynamic transition patterns from sequential recommendations in MetaTL, and meta-transitional learning facilitates fast learning for cold-start users with limited interaction. Meta-learning can improve sequential recommendations for cold-start users by inferring accurate sequential interactions.

    \item MAMO: Two memory matrices are used to store task-specific and feature-specific memories to support personalized parameter initialization and fast user preference prediction. 
    
    \item MetaCF: Discusses the cold-start problem in Collaborative Filtering (CF), where limited data is available for new users in the system. Previous approaches use user profiles, but these are not always available due to privacy concerns. MetaCF is a novel learning paradigm that leverages meta-learning to enable fast adaptation for new users. MetaCF learns a suitable initialization model for rapidly adapting to a new user. Adaptation rates are optimized in a fine-grained manner using Dynamic Subgraph Sampling to account for the dynamic arrival of new users. The proposed framework outperforms state-of-the-art baselines by a large margin in the cold-start scenario with limited user-item interactions.

\end{itemize}

\subsubsection{Evaluation Metrics}


Each user was tested on only one positive and true item during the experiment. Based on the predicted scores, observed edges were taken as positive samples for the user. 100 items without interaction with the user were randomly selected as negative samples. This method is commonly used in other works. The list of 100 negative and positive items was ranked, and Hit Ratio at rank 10 (HR@10) was applied as the evaluation metric to measure ranking performance. Mean Reciprocal Rank (MRR) was used to indicate the ranking of positive items, and Hit Rate (Hit) was evaluated for the top-1 prediction. If the positive item was ranked top-1, Hit@1 was equal to 1; otherwise, it was 0. It should be noted that HR@1 is equivalent to recall or NDCG for top-1 prediction.

\subsection{Overall Performance}

\begin{table*}[!t]
\centering
\caption{Experimental results of different methods under K=3 on three data sets}
\label{table:evaltable}
\begin{tblr}{
  cells = {c},
  cell{1}{1} = {r=2}{},
  cell{1}{2} = {c=3}{},
  cell{1}{5} = {c=3}{},
  cell{1}{8} = {c=3}{},
  hline{1,3-13} = {-}{},
  hline{2} = {2-11}{},
}
Methods            & Electronics    &                &                & Movies         &                &                & Beauty         &                &                \\
                   & $MRR$          & $Hit@1$        & $NDCG@5$       & $MRR$          & $Hit@1$        & $NDCG@5$       & $MRR$          & $Hit@1$        & $NDCG@5$       \\
BERT4Rec           & $0.323$        & $0.200$        & $0.319$        & $0.421$        & $0.220$        & $0.357$        & $0.341$        & $0.214$        & $0.338$        \\
MeLU               & $0.243$        & $0.136$        & $0.265$        & $0.336$        & $0.168$        & $0.302$        & $0.279$        & $0.160$        & $0.292$        \\
MAMO               & $0.296$        & $0.127$        & $0.313$        & $0.384$        & $0.194$        & $0.345$        & $0.310$        & $0.195$        & $0.336$        \\
MetaTL             & $0.320$        & $0.241$        & $0.324$        & $0.438$        & $0.319$        & $0.412$        & $0.328$        & $0.231$        & $0.335$        \\
MetaCF             & $0.330$        & $0.210$        & $0.313$        & $0.474$        & $0.276$        & $0.397$        & $0.340$        & $0.220$        & $0.322$        \\
\textbf{ClusterSeq} & $\mathbf{0.383}$ & $\mathbf{0.262}$ & $\mathbf{0.391}$ & $\mathbf{0.660}$ & $\mathbf{0.542}$ & $\mathbf{0.685}$ & $\mathbf{0.443}$ & $\mathbf{0.254}$ & $\mathbf{0.341}$ \\

\end{tblr}
\end{table*}

In this study, we evaluated the performance of ClusterSeq and state-of-the-art models under K = 3 on several datasets. The results are presented in Table \ref{table:evaltable}. The best-performing method in each column is highlighted in bold. The findings indicate that ClusterSeq outperforms the competing models in all datasets, demonstrating its effectiveness in providing accurate recommendations for cold-start users with limited interactions.

We started with basic neural models for sequential recommendations. We discovered that BERT4Rec performed poorly due to its inability to capture patterns in user interaction sequences and learn effective embeddings for cold-start users. However, utilizing transformers to extract sequential patterns proved more effective as they aggregate items with attention scores. This leads to more informative representations for users with limited interactions.

MeLU, MAMO, MetaCF, and MetaTL are meta-learning-based methods that provide cold-start recommendations. As MeLU and MAMO require side information about users and items, we used their historical interactions as side information. However, MeLU and MAMO failed to produce satisfying results, as they are designed for scenarios with abundant auxiliary user/item information, which is not the case here. On the other hand, MetaCF and MetaTL performed well in the sequential recommendation, highlighting the importance of fast adaptation in cold-start scenarios. Nevertheless, they still fell short of ClusterSeq's proposed clustering patterns for a cold-start sequential recommendation.

\subsection{Ablation Study}

We compare the proposed model with its variants and some baselines under different K values (i.e., how many interactions are initially present) to evaluate its effectiveness. Our original experiment demonstrated that BERT4Rec is the state-of-the-art sequential recommendation method, and MetaTL is one of the strongest cold-start baselines (and illustrates meta-transitional learning). Despite its high prediction power, BERT4Rec performs poorly on cold-start sequential recommendation tasks with a limited number of items. In sequential and cold-start user recommendations with different numbers of initial interactions, the proposed model can outperform state-of-the-art methods due to the well-designed optimization steps and clustering of users within the graph.

\begin{figure}[h]
\centering
\includegraphics[width=0.8\linewidth]{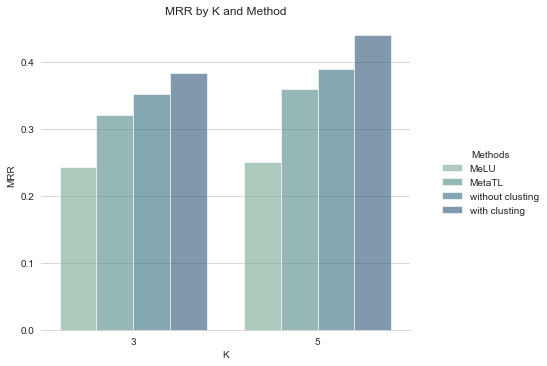}
\caption{Compare our model with its variant and some of the baselines under
different K values}
\label{fig:ablation_results}
\end{figure}

We compare the performance of our entire model (with the clustering module) to several model variants that do not include the clustering module. We evaluate these models on a standard benchmark dataset and report the results regarding our evaluation metrics.
Figure \ref{fig:ablation_results} shows the results of the ablation study. Clearly, the entire model achieves the highest evaluation metrics, indicating that the clustering module is necessary for the model to achieve its best performance. Performance is significantly reduced when the clustering module is removed.

\subsection{Parameter Analysis}
In this section, we investigate the impact of model parameters on the recommendation performance of our proposed model under cold-start scenarios. We examined how cluster number affects performance, followed by the impact of dimensions of user representations and learning rates.

\subsubsection{Number of clusters}
\begin{figure}[h]
\centering
\includegraphics[width=0.8\linewidth]{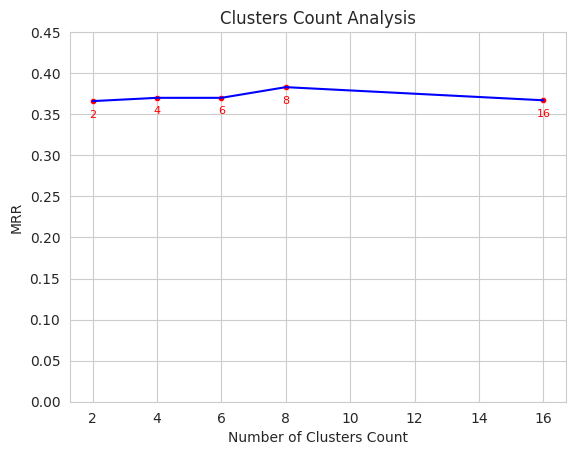}
\caption{Impact of number of clusters on recommendation performance}
\label{fig:num_clusters}
\end{figure}

To study the effect of the number of clusters, we vary the number of clusters and plot the performance of the proposed method in terms of MRR in Figure \ref{fig:num_clusters}. We observed that the performance of the proposed method is generally stable for different number of clusters. In particular, we find that the proposed model is robust against this hyperparameter which is hard to estimate.

\subsubsection{Impact of Embedding Dimensions}
\begin{figure}[h]
\centering
\includegraphics[width=0.8\linewidth]{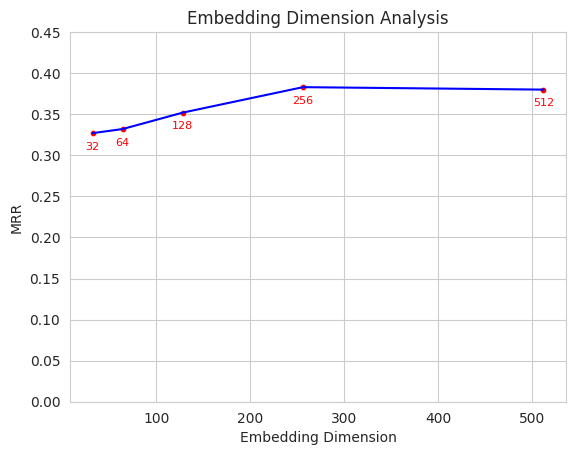}
\caption{Impact of embedding dimension on recommendation performance}
\label{fig:embedding_dim}
\end{figure}

Next, we explore the influence of the dimension of user embeddings on the recommendation performance of the proposed model. We vary the embedding sizes from 32 to 512 and plot the resulting performance in terms of MRR in Figure \ref{fig:embedding_dim}. Our model achieves optimal performance when the embedding dimension is set to 256. Our model is not generally stable around the optimal setting, indicating that it is important to set the embedding dimensions carefully.

\subsubsection{Impact of Batch size}
\begin{figure}[h]
\centering
\includegraphics[width=0.8\linewidth]{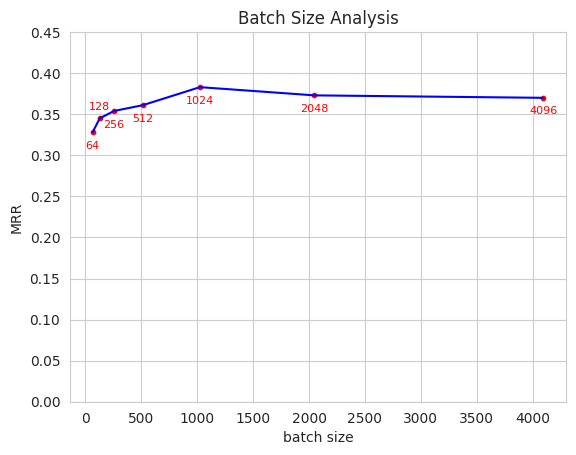}
\caption{Impact of batch size on recommendation performance}
\label{fig:batch_size}
\end{figure}

Lastly, we analyzed the impact of batch size on the performance of our model. We conducted experiments by varying the batch size from 64 to 4096 and evaluated the resulting performance in terms of MRR, as presented in Figure \ref{fig:batch_size}. Our observations show that the optimal performance of our model is achieved when the batch size is set to 1024. Additionally, we note that the training process can converge even with smaller batch sizes like 512 or 256. Overall, this analysis highlights the importance of selecting an appropriate batch size to achieve high performance in recommendation tasks.

To sum up, our analysis of the model parameters indicates that the recommendation performance of our model is stable for a reasonable range of hyperparameter values. However, some parameters are found to be critical for achieving optimal performance. Therefore, our findings emphasize the significance of meticulously tuning the model parameters to attain high recommendation performance in cold-start scenarios.

\section{Conclusion}

This paper proposes a method for extracting users' dynamic preferences through sequential personalized recommendations based on MAML. Meta-learning can adapt to different users, even with limited transactions, by formulating cold-start recommendations in a few-shot setting. Support adaptation is made in the training phase on a set of users with a few-shot transition in a sequence to mimic the targeted cold-start scenarios. Additionally, it avoids local optima, which is a substantial disadvantage of meta-learning in recommendation problems. Experiments have shown that the proposed method outperforms the current state-of-the-art methods in a cold-start scenario for three real-world datasets.

\bibliographystyle{ACM-Reference-Format}
\bibliography{sample-base}

\end{document}